\documentclass[fleqn,12pt,twoside]{article}
\usepackage{espcrc1}

\usepackage{graphicx}
\usepackage[figuresright]{rotating}

\newcommand{\AmS}{{\protect\the\textfont2
  A\kern-.1667em\lower.5ex\hbox{M}\kern-.125emS}}

\title{TWO-PHOTON EXCHANGE AND ELECTROMAGNETIC PROTON FORM FACTORS}

\author{Egle Tomasi-Gustafsson\address{DAPNIA/SPhN, CEA/Saclay, 91191 Gif-sur-Yvette Cedex,
France}}
        
\begin{document}

\maketitle

\begin{flushright}
{\it This talk is dedicated to the memory of Professor Michail P. REKALO}
\end{flushright}

\begin{abstract}
The presence of $2\gamma$-exchange in electron proton elastic scattering is discussed. From C-invariance and crossing symmetry, $2\gamma$ contribution induces a specific dependence of the reduced cross section on the variable $\epsilon$. No evidence of such dependence exists in the available experimental data.
\end{abstract}

\section{Introduction}

In 1999 M. P. Rekalo and myself investigated the possible presence of $2\gamma$ exchange \cite{Re99} in the precise data on the structure function $A(Q^2)$, obtained at the Jefferson Laboratory (JLab) in $electron-deuteron$ elastic scattering, up to a value of momentum transfer squared, $-t=Q^2=6$ GeV$^2$ \cite{Al99,Ab99}. A prescription for the differential cross section was derived from general properties of the hadron electromagnetic interaction, as the C-invariance and the crossing symmetry. The discrepancy from the two experiments \cite{Al99,Ab99}, which had been performed in different kinematical conditions, could not be explained in terms of a $2\gamma$ contribution, but the possibility of $2\gamma$-corrections was not excluded, starting from $Q^2=1$ GeV$^2$ and the necessity of dedicated experiments was pointed out \cite{Re99}.

The relative contribution of $2\gamma$ exchange, through its interference with the main (i.e., one-photon) mechanism is expected to be of the order of the fine structure constant, $\alpha={e^2}/{4\pi}\simeq {1}/{137}$. But, more than 30 years ago, it was observed 
\cite{Gu73} that the relative role of two-photon exchange can essentially increase in the region of high momentum transfer, due to the steep 
decreasing of the form factors (FFs). This effect can be observed in particular in $ed$-elastic scattering where it would appear already at momentum transfer squared of the order of 1 GeV$^2$. 

The main consequence of the presence of $2\gamma$ exchange is that the traditional description of the electron-hadron interaction in terms of electromagnetic currents (and electromagnetic FFs) can become incorrect. In one-photon exchange, two real amplitudes (functions of one variable, $Q^2$) fully describe elastic $ep$ scattering. If the $2\gamma$ exchange is present, one has to deal with three complex amplitudes, which are functions of two kinematical variables, $Q^2$ and  the polarization of the virtual photon, $\epsilon$. 

The presence of $2\gamma$ exchange leads to very complicated analysis of polarization effects. It destroys the linearity in the variable $\epsilon$ of the differential cross section for elastic $eN$ scattering \cite{Ro50}, and the relatively simple dependence of the ratio $P_x/P_z$ (the components of the final nucleon polarization in the scattering of longitudinally polarized electrons by an unpolarized nucleon target) on the ratio of the electric and magnetic FFs, $G_E/G_M$, which holds for the one-photon mechanism \cite{Re68}. It can be shown that the situation is not so involved, and that even in case of two-photon exchange, one can still use the formalism of FFs, if one takes into account the C-invariance of the electromagnetic interaction of hadrons \cite{Re04}. However, in this case only either a specific combination of three T-odd (or five T-even) polarization observables or measurements with positron and electron beams in the same kinematical conditions allow a model independent determination of FFs.

The exact calculation of the $2\gamma$-contribution to the amplitude of the $e^{\pm} p\to e^{\pm} p$-process requires the knowledge of the matrix element for the double virtual Compton scattering, $\gamma^*+N\to\gamma^*+N$, in a large kinematical region of colliding energy and virtuality of both photons, and can not be done in a model independent form. Therefore we follow another approach: general properties of the hadron electromagnetic interaction, as the C-invariance and the crossing symmetry, give rigorous prescriptions for different observables for the elastic scattering of electrons and positrons by nucleons, in particular for the differential cross section and for the proton polarization, induced by polarized electrons. These concrete prescriptions help in identifying a possible manifestation of the two-photon exchange mechanism and to avoid unjustified assumptions.  For example, symmetry properties appear in the spin structure of the amplitudes, with respect to the change $x\to -x$ with $x=\sqrt{(1+\epsilon)/(1-\epsilon)}$.
\section{Model independent analysis of experimental data}
Crossing symmetry allows to connect the matrix elements for the cross-channels: $e^-+N\to e^-+N$, in $s$--channel, and $e^++e^-\to N+\overline{N}$, in $t$--channel. 

The C-invariance of the electromagnetic hadron interaction and the corresponding selection rules can be  applied to the annihilation channel and this allows to find specific properties for one and two photon exchanges. 

To illustrate this, let us consider firstly the one-photon mechanism for $e^++e^-\to p+\overline{p}$. The conservation of the total angular momentum,  ${\cal J}$, allows one value, ${\cal J}=1$, and the quantum numbers of the photon: ${\cal J}^P=1^-$, $C=-1$. The selection rules with respect to C and P-invariances allow two states for $e^+e^-$ (and $p\overline{p}$): $S=1,~\ell=0 \mbox{~and~} S=1,~\ell=2\mbox{~with~} {\cal J}^P=1^-,$
where $S$ is the total spin and $\ell$ is the orbital angular momentum. As a result, the $\theta$-dependence of the cross section for $e^++e^-\to p+\overline{p}$, in the one-photon exchange mechanism is:
$
d\sigma/d \Omega\simeq a(t)+b(t)\cos^2\theta, 
$
where $a(t)$ and $b(t)$ are definite quadratic contributions of $G_E(t)$ and 
$G_M(t)$, $a(t),~b(t)\ge 0$ at $t\ge 4m^2$.

Let us consider now the $\cos\theta$-dependence of the $1\gamma\bigotimes 2\gamma$-interference contribution to the differential cross section of  $e^++e^-\to p+\overline{p}$. The spin and parity of the $2\gamma$-states 
is not fixed, in general, but only a positive value of C-parity, $C(2\gamma)=+1$, is allowed.
An infinite number of  states with different quantum numbers can contribute, and their relative role is determined by the dynamics of the process $\gamma^*+\gamma^*\to  p+\overline{p}$, with both virtual photons.

But the $\cos\theta$-dependence of the contribution to the differential cross section for the $1\gamma\bigotimes 2\gamma$-interference has a C-odd nature:
\begin{equation}
\displaystyle\frac{d\sigma}{d \Omega}^{(int)}(e^++e^-\to p+\overline{p})=\cos\theta[c_0(t)+c_1(t)\cos^2\theta+c_2(t)\cos^4\theta+...],
\label{eq:sig3}
\end{equation}
where $c_i(t)$, $i=0,1..$ are real coefficients, which are functions of $t$,  only. The following relation between kinematical variables in the crossing channels holds:  $
\cos^2\theta=(1+\epsilon )/(1-\epsilon)$. This odd $\cos\theta$ (or $x$)-dependence is essentially different from the even $\cos\theta$-dependence of the cross section for the one-photon approximation. It is, therefore,  incorrect to approximate the interference contribution to the differential cross section
(\ref{eq:sig3}) by a linear function in $\cos^2\theta$, because it is in contradiction with the C-invariance of hadronic electromagnetic interaction.

It follows that, in presence of $2\gamma$ exchange, the reduced elastic $ep$ cross section can be rewritten in the following general form:
\begin{equation}
\sigma_{red}(Q^2,\epsilon)= 
\epsilon G_E^2(Q^2)+\tau G_M^2(Q^2) +
\alpha F(Q^2,\epsilon), 
\label{eq:sred2}
\end{equation}
where $F(Q^2,\epsilon)$ is a real function describing the effects of the $1\gamma\bigotimes 2\gamma$ interference. In order to estimate the upper limit for a possible $2\gamma$ contribution to the differential cross section and the corresponding changing to $G_{E,M}(Q^2)$, we analyzed four sets of data \cite{Wa94,An94,Ch04,Ar04}, applying Eq. (\ref{eq:sred2}) with the following parametrization for $F(Q^2,\epsilon)$:
\begin{equation}
F(Q^2,\epsilon)\to \epsilon 
\sqrt{\displaystyle\frac{1+\epsilon}{1-\epsilon}}f(Q^2),~f(Q^2)=\displaystyle\frac{C}{(1+ Q^2\mbox{[GeV] }^2/0.71)^2(1+ Q^2\mbox{[GeV]}^2/m^2)^2},
\label{eq:sfit}
\end{equation}

\begin{figure}[h]
\vskip -1.3 true cm
\includegraphics[width=10cm]{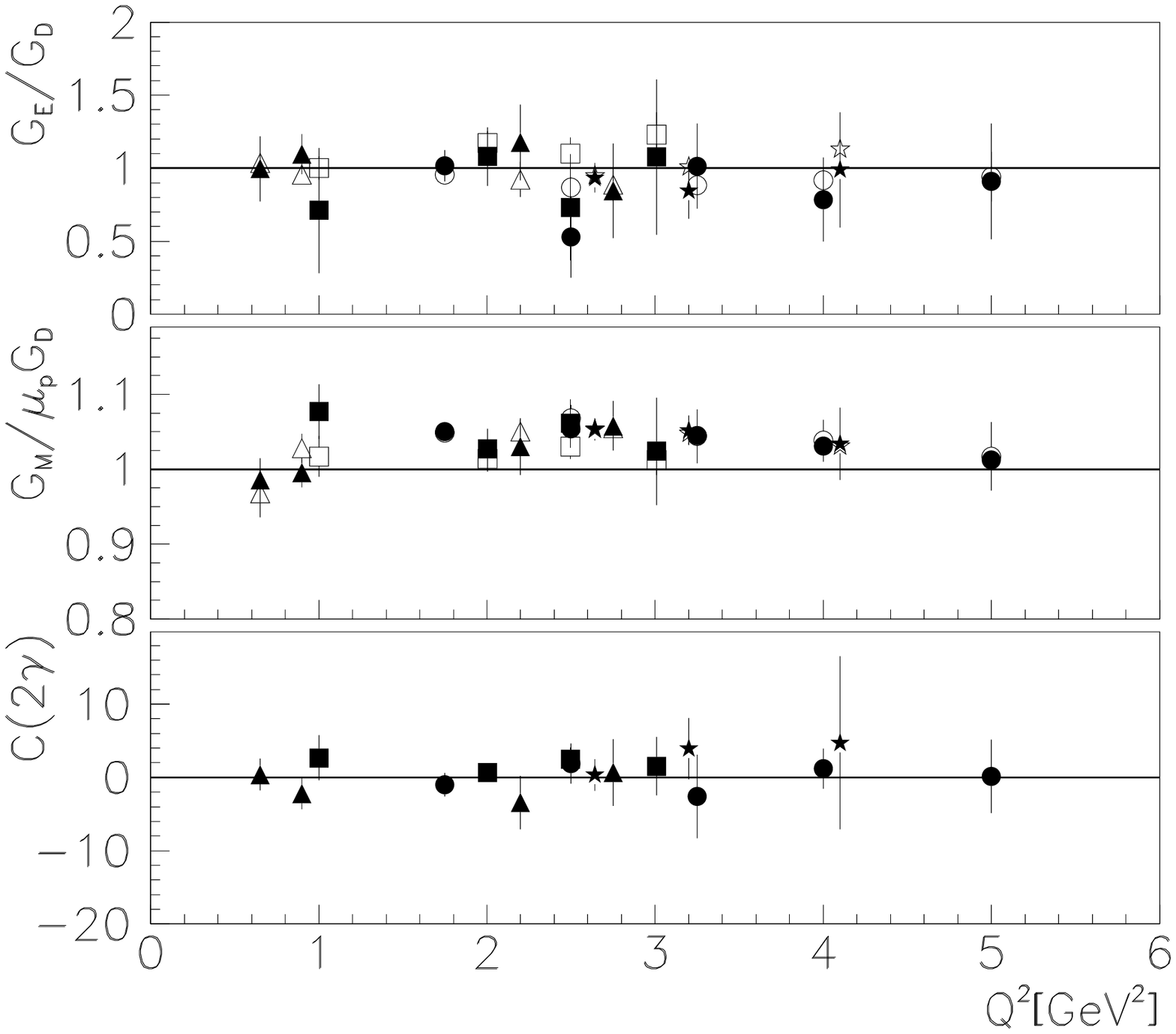}
\end{figure}

\vskip -.7 true cm
\begin{minipage}[b]{9 cm}
Figure 1. From top to bottom: $G_{E}/G_D$, $G_{M}/\mu G_D$  and the  two-photon contribution, $C$. $\mu=2.79$ is the proton magnetic moment, $G_D=(1+Q^2 [\mbox{GeV}^2]/0.71)^{-2}$. Published data are shown as open symbols, the present results which include  the $2\gamma$ contribution as solid symbols.
\end{minipage}

\begin{flushright}
\vspace*{-12. true cm}
\begin{minipage}[b]{6 cm}
where $C$ is a fitting parameter, $m$ is the mass of a tensor or vector meson with positive C-parity.
For $m\simeq 1.5$ GeV, one can predict that the relative role of the $2\gamma$ contribution should increase with $Q^2$. 
It is important to stress that Eq. (\ref{eq:sfit}) is a simple expression which contains the necessary symmetry properties of the $1\gamma\bigotimes 2\gamma$ interference, through a specific (and non linear) $\epsilon$ dependence.
Therefore, in presence of $2\gamma$, the dependence of the reduced cross section on $\epsilon$ can be parametrized as a function of three parameters, $G_E^2$, $G_M^2$ and $C$, according to Eqs. (\ref{eq:sred2}) and (\ref{eq:sfit}). In Fig. 1, from top to bottom, the electric and magnetic FFs, as well as the two photon parameter $C$, are shown as a function of $Q^2$ (solid symbols).
\end{minipage}
\end{flushright}

The previously published data, derived from the traditional Rosenbluth fits, are also shown (open symbols). Including a third fitting parameter, $C$, increases  the errors on the extracted FFs. The resulting parameter $C$ is compatible with zero.

\section{Conclusions}

From the present analysis it appears that the available data on $ep$ elastic scattering does not show any evidence of deviation from the linearity of the Rosenbluth fit, and hence of the presence of the $2\gamma$ contribution, when parametrized according to Eq. (\ref{eq:sfit}). 

Besides the deviation from the linearity of the Rosenbluth fit, other possible methods to test the presence of $2\gamma$-exchange in elastic electron hadron scattering can be listed: comparison of the cross section for scattering of unpolarized electrons and 
positrons (by protons or deuterons) in the same kinematical conditions;  specific polarization phenomena such as the appearance of T-odd polarization observables;  violation of definite relations between T-even polarization observables and structure functions \cite{Re99,Re04}.

The experimental evidence of the presence of the $2\gamma$-exchange and its quantitative estimation is very important. If this effect appears in elastic $ep$ scattering already in the range of momentum transfer investigated at JLab, the findings based on the one-photon assumption, will have to be reanalyzed at the light of a new and complicated formalism. In this case, most of the advantages related to the electromagnetic probe would be lost, as it was indicated already long ago \cite{Gu73}.
\section{Acknowledgments}

The results quoted here would not have been obtained wihout a fruitful collaboration and enlightning discussions with Professor M. P. Rekalo.

Thanks are due to G.I. Gakh for a careful reading of the manuscript.

\end{document}